\documentclass[lettersize,journal]{IEEEtran}

\IEEEoverridecommandlockouts

\usepackage[caption=false,font=normalsize,labelfont=sf,textfont=sf]{subfig}
\usepackage{textcomp}
\usepackage{stfloats}
\usepackage{verbatim}
\usepackage{cite}
\usepackage{amsmath,amssymb,amsfonts}
\usepackage{algorithmic}
\usepackage{graphicx,color}
\usepackage{float}
\usepackage{textcomp}
\usepackage{xfrac}
\usepackage{bbm}
\usepackage{array}
\usepackage{rotating}
\usepackage{gensymb}
\usepackage{comment}
\usepackage[acronym]{glossaries}
\usepackage{multirow}
\usepackage{lscape}
\usepackage{makecell}
\usepackage[inline]{enumitem}

\usepackage{adjustbox}
\usepackage{tikz}
\usetikzlibrary{shapes,arrows}
\usetikzlibrary{positioning}
\usetikzlibrary{decorations.text}

\newcommand{\rect}[2]{\Pi\left(\frac{#1}{#2}\right)}

\newcommand{\PRI}{\text{PRI}}
\newcommand{\CPI}{\text{CPI}}

\newcommand{\B}{\text{B}}
\newcommand{\D}{\text{D}}

\newcommand{\esp}[2]{\mathbb{E}_{#1}\left[#2\right]}

\renewcommand{\exp}[1]{e^{#1}}

\newcommand{\EbNo}{\sfrac{E_\text{b}}{N_0}}

\newcommand{\R}{\text{R}}
\newcommand{\C}{\text{C}}

\def\BibTeX{{\rm B\kern-.05em{\sc i\kern-.025em b}\kern-.08em
T\kern-.1667em\lower.7ex\hbox{E}\kern-.125emX}}

\begin{document}

\title{Interference Cancellation Schemes for Full-Duplex Integrated Sensing and Communication}

\author{\IEEEauthorblockN{François De Saint Moulin, Claude Oestges, Luc Vandendorpe\thanks{François De Saint Moulin is a Research Fellow of the Fonds de la Recherche Scientifique - FNRS.}}\\
\IEEEauthorblockA{ICTEAM, UCLouvain - Louvain-la-Neuve, Belgium\\}
e-mail: firstname.lastname@uclouvain.be}

\maketitle

\thispagestyle{plain}
\pagestyle{plain}


\begin{abstract}
In this paper, the introduction of interference cancellation in full-duplex joint radar and communication receivers is analysed. More specifically, a focus is made on scenarios in which the receiver simultaneously receives radar echoes from the environment, and communication signals from other joint radar and communication transceivers. First, the phase-coded frequency modulated continuous wave waveform designed for integrated sensing and communication is presented. Then, simple structures in which interference cancellation is only applied at the radar or communication function are proposed, relying on the information gathered at the other function. The detection probability and bit error rate improvements are analysed numerically w.r.t. different system parameters, such as the communication constellation, or the number of transmitted pulses. The introduction of error correcting codes is also considered. Next, iterative interference cancellation structures are investigated. Thanks to multiple interference cancellation layers, both the radar and communication performance are improved, and the robustness of the system to any scenario is enhanced. This is shown numerically through the analysis of the performance achieved by simple and iterative structures, which are compared for different radar echo to communication signal power ratio. Finally, a dynamic automotive scenario is considered. Leveraging on previous radar measurements, the parameters of the radar echo are inferred, and it is reconstructed beforehand, enabling to remove the first radar processing block. The complexity of the iterative structures is thus reduced, at the price of a slight performance reduction. A dynamic automotive scenario is considered, highlighting the impact of the tracking of the next vehicle ahead on the system.
\end{abstract}

\begin{IEEEkeywords}
ISAC, DFRC, full-duplex, interference cancellation, automotive scenario.
\end{IEEEkeywords}


\section{Introduction}
In recent years, with the spectrum scarcity and the convergence of the carrier frequency ranges used by radar and communication systems, the possibility to operate both functions using the same resources has received increased attention. Joint radar-communication systems perform both functions simultaneously with a single hardware platform \cite{Zheng2019,Liu2020,Luong2021,08999605,09737357}. Two approaches have been adopted: coexistence and co-design. In the former, different waveforms are used for each function, and interference between both functions must be mitigated. In the latter, a single waveform is transmitted, and trade-offs are made in the waveform design between both functions. These are also often called Dual Function Radar-Communication (DFRC), or Integrated Sensing and Communication (ISAC) systems. Numerous systems have been developed based on existing radar waveforms or communication protocols. Among these systems, as a non-exhaustive list, popular waveforms are the frequency hopping-based waveforms, chirp-based waveforms, e.g. Frequency Modulated Continuous Wave (FMCW), or Orthogonal Frequency Division Multiplexing (OFDM) waveforms.

\subsection{Related Works}
\subsubsection{FMCW ISAC Waveform}
In particular, the FMCW waveform is a popular radar waveform, widely used in automotive applications thanks to its low complexity and implementation cost. In order to introduce communication capabilities, the good cross-correlation properties of the up- and down-chirps enable to distinguish the transmission of a '0' or a '1', without affecting the radar function. This approach is adopted in \cite{Mietzner2020} in association with Alamouti space-time coding. However, the achievable rate is limited by the Pulse Repetition Interval (PRI) with inter-pulse coding. The same drawback applies to \cite{10289278}, which uses frequency shift keying to modify the starting frequency of the chirps for communication. In order to integrate multiple symbols during the pulse duration, \cite{Barrenechea2007} has proposed to modulate in amplitude the FMCW waveform. This has been extended for higher order modulations in \cite{Alabd2019,Uysal2020,Barreto2021} with different designs, and also generalised for a MIMO configuration in \cite{Bekar2021}. Hardware implementations have been demonstrated in \cite{McCormick2019,Uysal2020_2,Kumbul2021}. Performance comparisons are made in \cite{09769268,10172426} for different Phase-Coded FMCW (PC-FMCW) ISAC receivers. This waveform has been further extended by \cite{Ma2020,Ma2021} in a multi-carrier multi-antenna design, leveraging on index modulation. In order to limit distortions induced by phase changes at the power amplifier, Continuous Phase Modulations (CPM) and Minimum Shift Keying (MSK) are introduced to fulfil the communication function in \cite{Dou2017,Lampel2019}, and an experimental setup is shown in \cite{Bao2021}. This has also been used in \cite{Zhang2017} with a MIMO configuration for multiple users. Another issue of the PC-FMCW waveform is the bandwidth increase compared to the chirp bandwidth, induced by the introduction of the communication symbols. This problem is overcome in \cite{Zhang2017_2}, which separates the chirp in a three-section waveform to limit the excess bandwidth. To increase the data rate, the CPM FMCW ISAC waveform is associated with Low Density Parity Check (LDPC) codes in \cite{Li2019}. Another solution proposed by \cite{Ni2019} which avoids bandwidth increases is to modify the rate of transmission during the chirp duration. More recently, to further increase the data rate while keeping a low complexity at the radar receiver, \cite{09977883} has combined Orthogonal Time-Frequency Space (OTFS) modulation with FMCW waveforms. It is also combined with OFDM in \cite{10333763}, using the FMCW waveforms to simultaneously perform for radar sensing and channel estimation, acting like pilot signals for the communication function.

\subsubsection{Interference in Automotive Scenarios}
In automotive applications, interference between multiple radar or radar-communication systems remains a major issue owing to the constantly increasing number of vehicles (and consequently, of radar or radar-communication systems). This challenge has been tackled by the MOre Safety for All by Radar Interference Mitigation (MOSARIM) project \cite{MOSARIM}. Different interference mitigation techniques have been developed involving space, time, frequency, code, and polarisation domains. For instance, for coexisting radar systems, \cite{Rameez2018} has proposed an adaptive digital beamforming technique where the beamforming weights are updated depending on the estimated interference level. Modifications of chirp bandwidth, duration, and pulse repetition interval are implemented in \cite{Slavik2019,Kitsukawa2019,Kim2016}. Carrier sensing multiple access schemes are adopted in \cite{Ishikawa2019,Aydogdu2019}. Different multiple access strategies are also compared in \cite{Jin2021}. However, even if dynamic modifications of parameters are possible with radar systems, such solutions are not suited for ISAC systems since communication receivers are not collocated with the transmitter, except when the modification simultaneously carries information and helps for interference mitigation \cite{10333665}. Another approach applicable to both radar and ISAC scenarios is to consider interference suppression algorithms. This can be implemented through iterative interference reduction \cite{Umehira2018}, subspace error correction \cite{Yang-Ping2019}, short time Fourier transform interpolation \cite{Neemat2019}, sparsity-based \cite{Correas-Serrano2019} or other reconstruction methods \cite{Liu2020_2}.\\

\subsection{Full-Duplex ISAC Scenario}

\begin{figure}
    \centering
    \includegraphics[width=\linewidth]{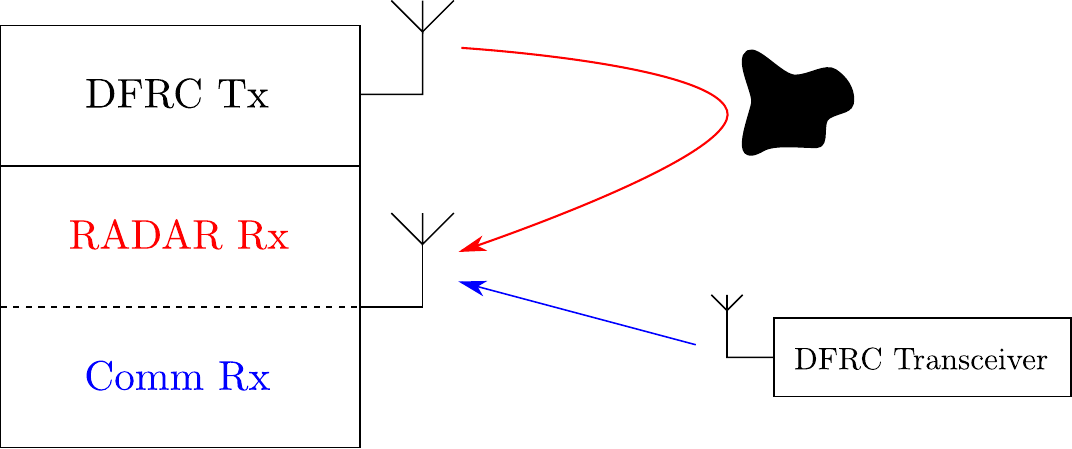}
    \caption{Full-duplex joint radar-communication scenario. The ISAC system simultaneously receives radar echoes from the targets, and communication signals from other systems.}
    \label{fig:FD_scenario}
\end{figure}

This paper focuses on full-duplex joint radar-communication scenarios, as illustrated in Fig. \ref{fig:FD_scenario}. Monostatic Single Input Single Ouptut (SISO) ISAC systems are considered. In such scenarios, a first issue to tackle is the self-interference from the transmitter to the receiver of the ISAC system. For instance, it is handled with successive interference cancellation or beamforming \cite{09764297,10159012}. Namely, an interference cancellation scheme for OFDM ISAC systems is proposed in \cite{6185152} in an automotive scenario. From an optimisation perspective, the full-duplex self-interference cancellation design is combined with a power allocation problem in \cite{09951144} to maximise both radar and communication performance.\smallskip

Assuming that the self-interference has been handled, the received signal at the ISAC system of interest can be written as 
\begin{equation}
    r(t) = r_\R(t) + r_\C(t) + w(t), \label{eq:full_duplex}
\end{equation}
where $w$ is an Additive White Gaussian Noise (AWGN). On the one hand, the signal $r_\R$ encompasses all the radar echoes from the targets. The $k^\text{th}$ echo is affected by a complex coefficient $\alpha_{\R k}$, a Doppler frequency $f_{\D\R k}$, and is delayed by a delay $\tau_{\R k}$. Denoting by $K$ the number of targets, and by $x_\R$ the transmitted ISAC waveform, it is expressed as 
\begin{equation}
    r_\R(t) = \sum_{k=0}^{K-1} \alpha_{\R k} \: \exp{j2\pi f_{\D\R k} t} \: x_\R(t-\tau_{\R k}). \label{eq:radar_signal}
\end{equation}
Note that, if there is some residual self-interference which does not induce saturation of the receiver front-end, it may be considered as a radar echo, and handled similarly. On the other hand, the signal $r_\C$ encompasses the uplink communication signals, arising from other ISAC systems. The $q^\text{th}$ communication signal $x_{\C q}$ is affected by a complex coefficient $\alpha_{\C q}$, a Doppler frequency $f_{\D\C q}$, and is delayed by a delay $\tau_{\C q}$. Denoting by $Q$ the number of communication signals, it is expressed as
\begin{equation}
    r_\C(t) = \sum_{q=0}^{Q-1} \alpha_{\C q} \: \exp{j2\pi f_{\D\C q} t} \: x_{\C q}(t-\tau_{\C q}). \label{eq:comm_signal}
\end{equation}
Note that multipath propagation can be handled by considering that several paths among the received communication signals are associated to the same transmitted signal $x_{\C q}$.\smallskip

Following \eqref{eq:full_duplex}, a second issue is the interference between the radar and communication functions. Since the ISAC system simultaneously receives communication signals from other systems while receiving radar echoes from the environment, the communication signals interfere with the radar function and the radar echoes interfere with the communication function. This paper aims to solve this second issue, i.e. to mitigate interference between both radar and communication functions in such scenarios using interference cancellation. Two interference cancellation structures, respectively at the radar or communication functions, were previously proposed in \cite{consanguinite}, performing well in scenarios where the radar echoes or the communication signal are respectively dominant. In this paper, our results from \cite{consanguinite} are extended by performing additional analyses, increasing the robustness of the proposed structures, and reducing the complexity.

\subsection{Contributions}
The contributions of this paper are summarised as follows:
\begin{itemize}
\item  Two joint radar-communication receiver chains integrating interference cancellation at the radar or communication function are presented. These receivers handle full-duplex radar-communication scenarios, in which an ISAC system receives simultaneously radar echoes and communication signals. These structures perform well in scenarios where the radar echoes or the communication signal is dominant. They are particularised with PC-FMCW ISAC systems, but similar received chain for any joint radar-communication waveforms (i.e. OFDM systems) can be designed following the proposed approach. The improvements of detection probability and Bit Error Rate (BER) thanks to interference cancellation are evaluated for multiple system parameters, such as the communication constellation, number of transmitted pulses, and zero-padding factor. Further improvement of the joint radar-communication receiver chain integrating interference cancellation at the radar function is performed thanks to the introduction of error correcting codes. Their impact on the inherent interference mitigation of the PC-FMCW ISAC waveform is also discussed.
\item Based on the presented receiver chains, an iterative structure integrating interference cancellation at both the radar and communication functions is proposed, in order to increase the robustness of the system for scenarios in which the communication signal or the radar echo is more powerful, or both are received at equivalent power levels. The performance of such structures is analysed w.r.t. the ratio between the power of the radar echo and communication signal, showing that the performance achieved by iterative structures are equivalent or better compared to the simple structures, whatever the scenario, at the price of a higher complexity.
\item In order to reduce the complexity of the iterative receiver, a last interference cancellation structure is proposed, leveraging on dynamic scenarios in which successive radar measurements are performed for targets tracking. The performance achieved by all structures are for instance evaluated and compared in a dynamic automotive scenario in which a vehicle simultaneously detects the next vehicle behind, while receiving a communication signal from the infrastructure.
\end{itemize}

\subsection{Structure of the Paper}
\begin{figure*}
\centering
\includegraphics[width=0.8\linewidth]{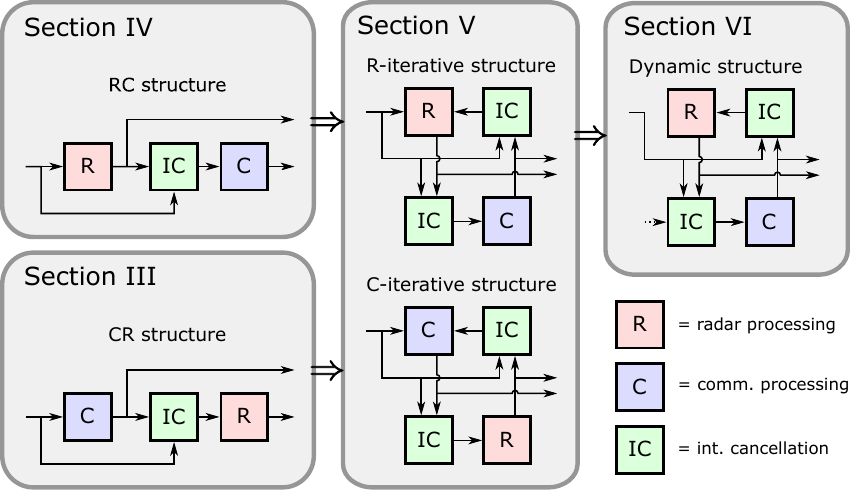}
\caption{Structure of this paper. The first solid arrow entering the blocks from the left represents the received signal. The solid arrows at the output of the radar and communication blocks represents the number of targets and their parameters, or the communication channel parameters and the decoded symbols. The solid arrows at the output of the interference cancellation blocks represents the received signal in which the radar or communication interference has been cancelled. The dotted arrow in the dynamic structure represents radar information forwarded from previous measurements.}
\label{fig:paper_structure}
\end{figure*}

First, Section \ref{sec:PC-FMCW} details the PC-FMCW DFRC processing.  Then, Sections \ref{sec:interference_RADAR} and \ref{sec:interference_comm} analyse the impact of correlated interference (systems with identical parameters), respectively on the radar and communication receivers. Additionally, the performance of interference cancellation on both functions is studied, in cooperation with the other function. Next, iterative structures are proposed in Section \ref{sec:iterative_IC} to further improve the performance and robustness of the system. Finally, the application of the iterative structure in dynamic scenarios is studied in Section \ref{sec:dynamic_IC}, and the possibility to leverage on previous radar measurement is analysed. The paper structure is summarised in Fig. \ref{fig:paper_structure}, with schematics of the different interference cancellation receivers that are developed in this work.

\section{PC-FMCW DFRC System}
\label{sec:PC-FMCW}
With an FMCW radar, multiple increasing or decreasing chirps are transmitted. Assuming that the transmission occurs in a frequency band $f \in [f_c - B/2,f_c + B/2]$ where $f_c$ is the carrier frequency and $B$ the bandwidth, if $P$ pulses of duration $T$ are transmitted with a PRI denoted as $T_\PRI$, the instantaneous frequency of the transmitted signal is given by
\begin{equation}
f(t) = f_c + \sum_{p=0}^{P-1} \left[-\frac{B}{2}+\frac{B}{T}\left(t-pT_\PRI\right)\right]\rect{t-pT_\PRI}{T},
\end{equation}
where $\Pi$ is the non-centred rectangular function, i.e. $\Pi(t)=1$ if $t\in[0,1]$. Note that $T_\PRI \geq T$, allowing the insertion of a guard interval $T_g = T_\PRI - T$ between successive pulses. The corresponding transmitted signal in baseband is given by
\begin{equation}
s_\R(t) = \sum_{p=0}^{P-1} \exp{-j\pi B \left(t-pT_\PRI\right)}\:\exp{j\pi\frac{B}{T}\left(t-pT_\PRI\right)^2} \: \rect{t-pT_\PRI}{T}. \label{eq:chirp_train}
\end{equation}
Let us consider a complex pulse modulation where symbols $\textbf{I} \in \mathbb{C}^{L_c \times P}$ are pulse shaped with a rectangular filter of width $T_c=T/L_c$, with $L_c$ the number of transmitted symbols per pulse. The communication payload is expressed in baseband as
\begin{equation}
s_\C(t) = \sum_{p=0}^{P-1}\sum_{l=0}^{L_c - 1} I(l,p) \: \rect{t-pT_\PRI - lT_c}{T_c}. \label{eq:comm_payload}
\end{equation}
In a PC-FMCW ISAC system, the communication payload of \eqref{eq:comm_payload} is modulated with the FMCW waveform of \eqref{eq:chirp_train}. The joint waveform is thus given by
\begin{align}
x(t) &= s_\R(t) \: s_\C(t) \\
\nonumber &=\sum_{p=0}^{P-1}\sum_{l=0}^{L_c - 1} I(l,p) \: \rect{t-pT_\PRI - lT_c}{T_c} \\
&\qquad\qquad\qquad\cdot\exp{-j\pi B \left(t-pT_\PRI\right)}\:\exp{j\pi\frac{B}{T}\left(t-pT_\PRI\right)^2}. \label{eq:PC_FMCW}
\end{align}
Let us define the notation in fast- and slow-time of a signal $x$ as $x(t,p) = x\left(t+pT_\PRI\right)$ with $t\in [0,T_\PRI]$ and $p=0,...,P-1$. This notation simplifies the joint waveform as follows:
\begin{equation}
    x(t,p) = \sum_{l=0}^{L_c - 1} I(l,p) \: \rect{t - lT_c}{T_c} \: \exp{-j\pi B t}\:\exp{j\pi\frac{B}{T}t^2}.
\end{equation}
In this paper, it is assumed that every ISAC system transmits a PC-FMCW signal. Thus, the signals $x_\R$ and $x_{\C q}$ of \eqref{eq:radar_signal} and \eqref{eq:comm_signal} are all expressed in the form of \eqref{eq:PC_FMCW}, and they all contain a radar and a communication component. Therefore, on the one hand at the radar function, the signal $x_\R$ is the signal of interest, and the radar processing aims to extract the information from the radar component of the joint waveform. On the other hand, at the communication function, the signals $x_{\C q}$ are the signals of interest, and the communication processing aims to extract the information from the communication component of the joint waveform.

\subsection{Processing at the Radar Function}
\label{sec:radar_processing}

Let us focus on the signal $r_\R$ of \eqref{eq:radar_signal}, encompassing all the radar echoes from the targets. In fast and slow-time notation, it can be developed as 
\begin{align}
 &r_\R(t,p) = \sum_{k=0}^{K-1} \alpha_{\R k}\:\exp{-j2\pi f_{\D\R k} T_\PRI p} \sum_{l=0}^{L_c - 1} I_\R(l,p) \\ \nonumber & \cdot \rect{t -\tau_{\R k} - lT_c}{T_c} \exp{j2\pi f_{\D\R k} t}\: \exp{-j\pi B \left(t-\tau_{\R k}\right)}\:\exp{j\pi\frac{B}{T}\left(t-\tau_{\R k}\right)^2},
\end{align}
where $\textbf{I}_\R \in \mathbb{C}^{L_c \times P}$ are known symbols transmitted by the ISAC system of interest. At the radar function, a dechirp operation is first performed by mixing the received with the conjugate of the modulating FMCW signal of \eqref{eq:chirp_train}. Assuming that $\max_{k} \left\{\tau_{\R k}\right\} \leq T_g$, or equivalently that all high frequencies components generated by targets with a larger delay than $T_g$ are filtered out, the signal at the output of the dechirp operation is given by
\begin{align}
\nonumber y_\R(t,p) &= \sum_{k=0}^{K-1} \beta_{\R k} \: \exp{j2\pi f_{\D\R k}T_\PRI p} \: \exp{-j2\pi f_{\B\R k} t} \\
& \qquad \cdot \sum_{l=0}^{L_c-1} I_\R(l,p) \: \rect{t -\tau_{\R k} - lT_c}{T_c},
\end{align}
where $\beta_{\R k} \triangleq \alpha_{\R k} \: \exp{j\pi B \tau_{\R k}} \: \exp{j \pi \frac{B}{T} \tau_{\R k}^2}$, and $f_{\B \R k} \triangleq \frac{B}{T} \tau_{\R k} - f_{\D \R k}$ is the beat frequency associated to the $k^\text{th}$ echo.\smallskip

Then, the communication part of the joint waveform must be compensated, otherwise the integration of communication symbols induces a dispersion of the targets power in the delay-Doppler map, as further analysed in Section \ref{sec:RADAR_int_theoretical_analysis}. The symbol pulses contained in the echo coming back from the $k^\text{th}$ target are delayed with a delay $\tau_{\R k}$, initially unknown at the receiver. Nonetheless, assuming that $f_{\B\R k}\approx -\frac{B}{T}\tau_{\R k}$, i.e. that the Doppler frequency of the echoes are negligible w.r.t. the generated beat frequencies, the beat frequency of each path is proportional to its delay. Therefore, the multiple copies of the communication component of the joint waveform are aligned with the start of the chirp using a Group Delay Filter (GDF) $H_{\text{gd}}$ \cite{09769268}. Since the $k^\text{th}$ target echo associated with a beat frequency $f_{\B\R k} \approx -\frac{B}{T}\tau_{\R k}$ should be delayed by a group delay $\tau_{\text{gd} k} = \frac{T}{B} f_{\B\R k}$, the filter is designed with modulus and phase given by 
\begin{equation}
|H_{\text{gd}}(f)|=1,\: \tau_{\text{gd}}(f) = \frac{T}{B}f \:\leftrightarrow\: \text{arg}\{H_{\text{gd}}(f)\} = - \pi \frac{T}{B} f^2,
\end{equation} 
where $\tau_\text{gd}$ is the group delay of the filter. If the envelope is slowly varying compared to the beat frequency, the signal at the output of the GDF expressed in pulse notation is approximatively given by
\begin{align}
\nonumber y_{\R,\text{gd}}(t,p) &\approx \sum_{k=0}^{K-1} \gamma_{\R k} \: \exp{j 2\pi f_{\D\R k} T_\PRI p} \: \exp{-j2\pi f_{\B\R k} t} \\ 
&\qquad \cdot \sum_{l=0}^{L_c - 1} I_\R(l,p)\:\rect{t-lT_c}{T_c},
\end{align}
where $\gamma_{\R k} \triangleq \beta_k \: \exp{-j\pi \frac{B}{T} \tau_{\R k}^2} = \alpha_{\R k} \: \exp{j\pi B \tau_{\R k}}$. After echoes alignment in time, the communication part can be compensated:
\begin{align}
\nonumber z_\R(t,p) &= \sum_{l'=0}^{L_c-1} y_{\R,\text{gd}}(t,p) \: \frac{I_\R^*(l',p)}{\left|I_\R(l',p)\right|^2} \: \rect{t-l'T_c}{T_c} \\
&\approx \sum_{k=0}^{K-1} \gamma_{\R k} \: \exp{j 2\pi f_{\D\R k} T_\PRI p} \: \exp{-j2\pi f_{\B\R k} t}.\label{eq:comm_compensation}
\end{align}
Finally, the delay-Doppler map is obtained after sampling and applying a 2D-Discrete Fourier Transform (DFT). This map is typically fed to a detector, e.g. a Constant False Alarm Rate (CFAR) detector, in order to deduce the number of targets and their parameters \cite{richards2005fundamentals}.

\subsection{Processing at the Communication Function}
\label{sec:comm_processing}

Let us focus on the signal $r_\C$ of \eqref{eq:comm_signal}, encompassing all the uplink communication symbols from other ISAC systems. In fast and slow-time notation, it can be developed as 
\begin{align}
    &r_\C(t,p) = \sum_{q=0}^{Q-1} \alpha_{\C q}\:\exp{-j2\pi f_{\D\C q} T_\PRI p} \sum_{l=0}^{L_c - 1} I_{\C q}(l,p) \\ \nonumber & \cdot \rect{t -\tau_{\C q} - lT_c}{T_c} \exp{j2\pi f_{\D\C q} t}\: \exp{-j\pi B \left(t-\tau_{\C q}\right)}\:\exp{j\pi\frac{B}{T}\left(t-\tau_{\C q}\right)^2},
\end{align}
where $\textbf{I}_{\C q} \in \mathbb{C}^{L_c \times P}$ are the communication symbols transmitted by the $q^\text{th}$ ISAC system. Let us first assume a transmission over a flat fading communication channel. Focusing on the $q^\text{th}$ ISAC system, a time synchronisation is performed to align the signal with the start of the $q^\text{th}$ uplink signal. Thus, the first pulse of the payload is assumed to be generated from pilot symbols, enabling to perform a correlation in time. Then, a dechirp operation is performed to demodulate the communication part of the ISAC waveform. The output of both operations is given by 
\begin{align}
\nonumber &y_{\C q}(t,p) = r_{\C}(t + \tau_{\C q},p) \: \exp{j\pi B t} \: \exp{-j\pi \frac{B}{T}t^2} \: \rect{t}{T} \\
&= \beta_{\C q p} \: \exp{j2\pi f_{\D\C q}t} \: \sum_{l=0}^{L_c-1} I_{\C q}(l,p) \:\rect{t - nT_c}{T_c} + \text{IUI},
\end{align}
where $\beta_{\C q p} \triangleq \alpha_{\C q} \: \exp{-j 2\pi f_{\D\C q} \tau_{\C q}} \: \exp{j 2\pi f_{\D\C q}T_\PRI p} \: \exp{j\pi B \tau_{\C q}}$, and $\text{IUI}$ denotes the interference from the other communicating ISAC systems. Finally, usual communication processing, namely matched filtering, carrier frequency offset correction, equalisation, and symbol decision, are applied to recover the symbols $\textbf{I}_{\C q}$. The carrier frequency offset is either estimated from the first pulse, assumed to be generated from pilot symbols, or from pilots symbols inserted at the start of successive pulses to increase the resolution of the estimation. For instance, possible techniques which may be applied are the Schmidl \& Cox algorithm \cite{schmidl1997robust}, or Fourier transform-based methods \cite{alabd2019partial}. Note that the same processing can also be applied for transmissions over frequency selective channels, but this is not considered in this work.

\subsection{Summary}

\begin{figure}
    \centering 
    \includegraphics[width=\linewidth]{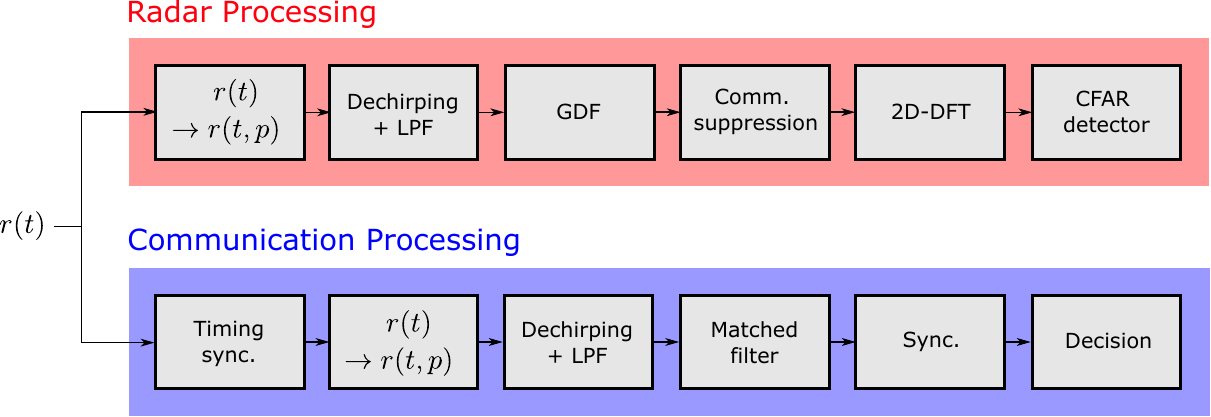}
    \caption{Phase-coded FMCW ISAC system.}
    \label{fig:FMCW_ISAC_chain}
\end{figure}

Figure \ref{fig:FMCW_ISAC_chain} summarises the processing of phase-coded FMCW systems. On the one hand, for the radar function, most complexity is introduced by the CFAR detector, as the noise power is estimated for each cell of the delay-Doppler map based on the power of the neighbouring cells. On the other hand, for the communication function, the synchronisation algorithms are the most complex operations, especially the timing synchronisation. \bigskip

In the following sections, scenarios with one radar and communication link are considered, i.e. $K=1$ and $Q=1$. On the one hand, when considering the interference between multiple communication links, interference cancellation has already been widely studied within the communication literature, e.g. for Non Orthogonal Multiple Access (NOMA) applications \cite{7676258,7263349}. On the other hand, the impact of the number of targets on the communication function has been evaluated in \cite{consanguinite}, without providing new insights compared to communication-only scenarios. Thus, this motivates to simplify the performance analysis of the interference cancellation schemes in a full-duplex ISAC scenario by considering only one radar and communication link, and highlighting the impact of the radar and communication parameters on the achieved performance. Additionally, this also assumes flat fading radar and communication links. This is motivated by the high frequencies considered in automotive radar applications \cite{ETSI2}, and for future wireless communication networks \cite{ETSI}. At such frequencies, the channels are usually assumed to be highly Ricean.

\section{Interference Cancellation at the Radar Function}
\label{sec:interference_RADAR}
Let us first focus on the radar function. The scenario presented in Fig. \ref{fig:FD_scenario} is considered, in which target echoes are received simultaneously with a communication signal coming from another system sharing the same system parameters ($f_c$, $T_\PRI$, $T_\CPI$, $B$, $T$, and $L_c$). At the radar function, the signals of interest are the radar echoes, and the communication signal emitted by another system is interfering.

\subsection{Theoretical Analysis}
\label{sec:RADAR_int_theoretical_analysis}

\begin{figure}
    \centering
    \includegraphics[width=0.9\linewidth]{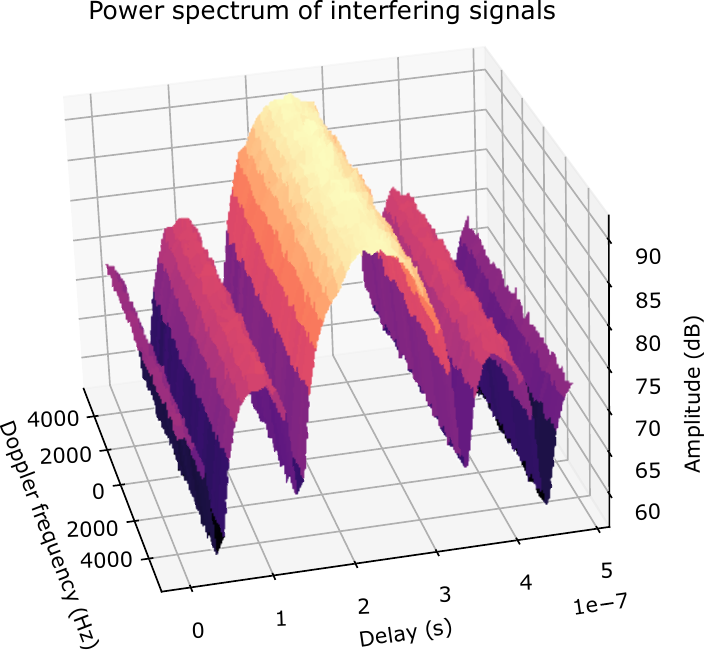}
    \caption{Illustration of the power dispersion of an interfering signal. The parameters are $\normalfont\alpha_{\C} = 1$, $\normalfont\tau_\C = 0.25 \: \mu\text{s}$, $\normalfont f_{\D\C} = -300\:\text{Hz}$, $P=50$, $N=10$, $M=2000$, $\normalfont T = 100 \:\mu\text{s}$, $\normalfont T_\PRI = 100.5 \:\mu\text{s}$, $\normalfont B = 100\:\text{MHz}$.}
    \label{fig:DD_map_int}
\end{figure}

The communication signal is filtered at the radar function after dechirping if the total Doppler frequency, namely the sum of the Doppler frequency of the communication channel and the beat frequency generated by the dechirping operation, is higher than the cut-off frequency of the receiver Low-Pass Filter (LPF). However, if the relative delay between the interferer and the receiver is lower than the maximum detectable delay of the radar receiver, the delay-Doppler map will be corrupted by the interfering communication signal. Following the radar processing presented in Section \ref{sec:comm_processing}, the communication signal which has been processed at the radar function is similar to \eqref{eq:comm_compensation}, except that the communication symbols have not been well compensated. In discrete time, it is written as 
\begin{align}
\nonumber z_{\C}(l,p) &= \gamma_{\C} \: \exp{j 2\pi f_{\D\C} T_\PRI p} \: \exp{-j 2\pi f_{\B\C} T_s l} \\&\quad \cdot  \sum_{n=0}^{ML_c-1} \frac{I_{\C}(n,p) I_{\R}^*(n,p)}{|I_{\R}(n,p)|^2} \: \rect{l-nM}{M}, \label{eq:interference_radar}
\end{align}
where $\gamma_\C = \alpha_\C \: \exp{j\pi B \tau_\C}$, and $T_s = T_c/M$ is the sample period. In order to evaluate the impact of this interference on the delay-Doppler map while accounting for the random complex coefficients and symbols, one may compute the expectation of the autocorrelation of $z_{\C}$ over the fast- and slow-time, and then apply Fourier transforms over both dimensions. The expectation of the autocorrelation of $z_{\C}$ is computed as 
\begin{align}
    \nonumber &\Gamma_{z_\C}(\Delta_l,\Delta_p) = \esp{}{\sum_{p=0}^{P-1}\sum_{n=0}^{ML_c-1} z_{\C}(l,p) z_{\C}^*(l-\Delta_l,p-\Delta_p)}\\
    &= \sigma_{\C}^2\: M L_c P\: \delta(\Delta_p) \: \exp{-j2\pi f_\text{BC} T_s \Delta_l} \: \Lambda\left(\frac{\Delta_l}{M}\right). \label{eq:interference_radar_autocorr}
\end{align}
with $\Lambda$ being the triangular function, i.e. $\Lambda(x) = \text{max}(1-|x|,0)$. This equality is obtained assuming independent symbols drawn from a Phase Shift Keying (PSK) constellation, such that 
\begin{equation}
\esp{}{I_\C(n,p) I_{\C}^*(n',p')} = \delta(n-n') \delta (p-p'). \label{eq:cond_symbols}
\end{equation}
Finally, applying Fourier transforms over both dimensions provides an insight of the power dispersion of the signal into the delay-Doppler domain:
\begin{equation}
\mathcal{P}_{x_\C}(\tau,f) = \sigma_{\C}^2\:P L_c \: \frac{\sin^2\left(\frac{\pi}{L_c}\left(\tau+f_\text{BC} T\right)\right)}{\sin^2\left(\frac{\pi}{M L_c}\left(\tau+f_\text{BC} T\right)\right)}, \label{eq:power_dispersion_int}
\end{equation} 
where $\sigma_\C^2 \triangleq \esp{}{|\gamma_\C|^2} = \esp{}{|\alpha_\C|^2}$. The dimensions $\tau$ and $f$ respectively refer to the Fourier transforms along the dimensions $\Delta_l$ and $\Delta_p$. This shows that the power of the communication signal is spread uniformly in the Doppler dimension, and follows a sinus over sinus shape in the delay dimension. The main lobe width is proportional to $L_c$. The maximum value of the power is located around the normalised beat frequency $f_{\B\C} T$, and is equal to $\sigma_\C^2 P L_c M^2$. This is illustrated in Fig. \ref{fig:DD_map_int}.\smallskip

This can be compared to the power distribution of a radar echo, in which the communication part has been well compensated. In that case, the processed echo is written in discrete time following \eqref{eq:comm_compensation} as
\begin{equation}
z_\R(l,p) = \gamma_\R\:\exp{j2\pi f_\D T_\PRI p}\:\exp{-j2\pi f_{\B \C} T_s l}.
\end{equation}
The expectation of the autocorrelation of $z_\R$ is computed as
\begin{align}
    \nonumber &\Gamma_{z_\R}(\Delta_l,\Delta_p) = \esp{}{\sum_{p=0}^{P-1}\sum_{n=0}^{ML_c-1} z_{\R}(l,p) z_{\R}^*(l-\Delta_l,p-\Delta_p)}\\
    &= \sigma_{\R}^2\: M L_c P \: \exp{j 2\pi f_{\D\R} T_\PRI \Delta_p}\: \exp{-j2\pi f_\text{BR} T_s \Delta_l}.
\end{align}
Applying Fourier transforms over both dimensions provides the power dispersion of the signal into the delay-Doppler domain:
\begin{align}
    & \mathcal{P}_{x_\R}(\tau,f) = \sigma_{\R}^2\:M L_c P\\
    \nonumber &\qquad\cdot  \frac{\sin\left(\pi\left(f-f_\text{DR} T_\CPI\right)\right)}{\sin\left(\frac{\pi}{P}\left(f - f_\text{DR} T_\CPI\right)\right)}\frac{\sin\left(\pi\left(\tau-f_\text{BR} T\right)\right)}{\sin\left(\frac{\pi}{M L_c}\left(\tau-f_\text{BR} T\right)\right)},
\end{align}
where $\sigma_\C^2 \triangleq \esp{}{|\gamma_\R|^2} = \esp{}{|\alpha_\R|^2}$, and $T_\CPI \triangleq T_\PRI P$ is the coherent processing interval. This shows that the radar echo power follows a sinus over sinus shape in both the delay and Doppler dimensions. The maximum value of the power is located around the normalised beat frequency $f_{\B\R} T$ and normalised Doppler frequency $f_{\D\R}T_\CPI$, and it is equal to $\sigma_\R^2 P^2 L_c^2 M^2$. Comparing this result to the maximum value of \eqref{eq:power_dispersion_int}, one observes that, for the same variances of the radar and communication channels, a Signal-to-Noise Ratio (SNR) gain of $10\log_{10} (PL_c)$ dB is obtained for the radar echoes compared to the interfering communication signals. Note that, if the radar echoes were not aligned with the GDF, the communication symbols would not be well compensated, and the power of the radar echoes would also be dispersed in the delay-Doppler map. This justifies the use of a GDF. It can be shown that similar results are obtained for OFDM ISAC systems, in which the symbols are not well compensated at the radar function.\smallskip

With a higher number of transmitted symbols per pulse $L_c$, the main lobe is widened, and nearly constant interference is observed on the whole delay-Doppler map. Nonetheless, the interfering power is spread and not suppressed, leading to an increased noise floor, which degrades severely the radar detection performance. This motivates the introduction of interference cancellation at the radar function.

\subsection{Interference Cancellation}
\begin{figure}
\centering
\includegraphics[width=\linewidth]{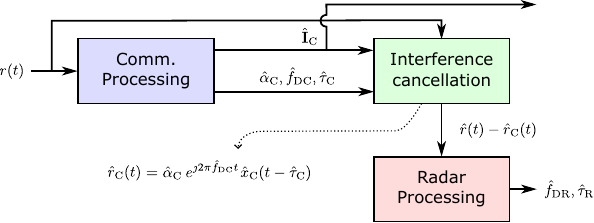}
\caption{Receiver chain implementing the CR structure. The communication function forwards estimates of the attenuation, phase shift, delay, Doppler frequency, and decoded symbols of the communication signal to the radar function.}
\label{fig:RADAR_IC_chain}
\end{figure}

In scenarios where the power of the communication signal is assumed to be larger than the power of the radar echo, interference cancellation can be applied at the radar function. Following the processing of Section \ref{sec:comm_processing}, from the received signal $r$, the communication signal $x_\C$ is first decoded at the communication function, providing an estimation of the communication symbols $\textbf{I}_\C$. Moreover, the synchronisation parameters, i.e. $\alpha_\C$, $f_{\D\C}$, and $\tau_{\C}$, are also estimated. All these parameters are forwarded to the interference cancellation block of the radar function, which generates an estimate of the signal $r_\C$: 
\begin{equation}
\hat{r}_{\C}(t) = \hat{\alpha}_\C \: \exp{j2\pi \hat{f}_{\D\C} t} \: \hat{x}_\C(t-\hat{\tau}_\C),
\end{equation}
where the symbol $\hat{\cdot}$ designates estimated quantities. Then, the radar processing is applied on the received signal $r$ from which the uplink communication signal has been suppressed: 
\begin{equation}
\tilde{r}_{\R}(t) = r(t) - \hat{r}_{\C}(t). 
\end{equation}
This receiver is named \textit{CR structure}, based on the order in which the radar and communication function are processed. It is illustrated in Fig. \ref{fig:RADAR_IC_chain}. Note that this process can be repeated for each communication signal if there is multiple ISAC systems communicating in uplink, or for multipath propagation. 

\begin{table}
\centering
\caption{Simulation parameters for Section \ref{sec:interference_RADAR}. The false alarm probability requirement is denoted by $\mathcal{P}_\text{FA}$.}
\begin{tabular}{|l|l|l|}
\hline
$B$ = 20 MHz & $T$ = 100$\:\mu$s & $T_g$ = 0.5 $\mu$s \\\hline
$P$ = 50 & $L_c$ = 100 & $\mathcal{P}_\text{FA} = 10^{-4}$ \\\hline
$|\alpha_{\R}|$ = 0.1 & $\tau_{\R}$ = 0.1 $\mu$s & $f_{\D\R}$ = 1000 Hz \\\hline
$|\alpha_\C|$ = 1 & $\tau_\C$ = 0.25 $\mu$s & $f_{\D\C}$ = -300 Hz \\\hline
\end{tabular}
\label{tab:scenar1}
\end{table}

\begin{figure}
    \centering
    \includegraphics[width=\linewidth]{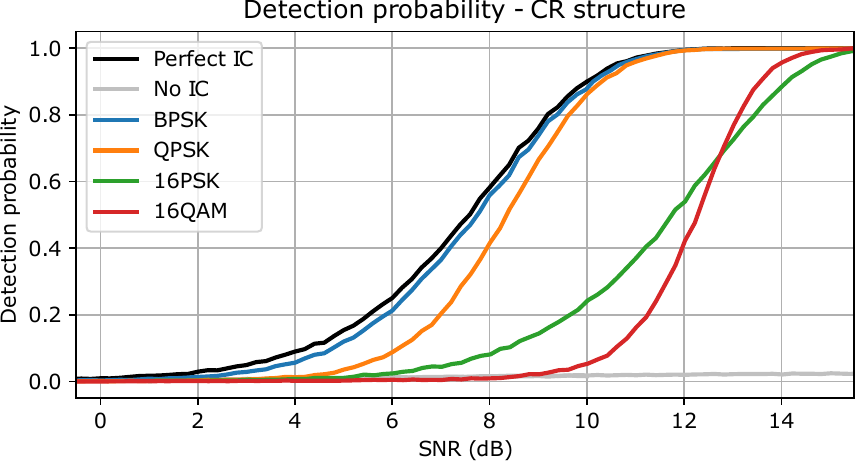}
    \caption{Detection probability against SNR with BPSK, QPSK, 16-PSK and 16-QAM modulation, with and without interference cancellation.}
    \label{fig:RIC_P_D}
    \vspace*{0.2cm}

    \includegraphics[width=\linewidth]{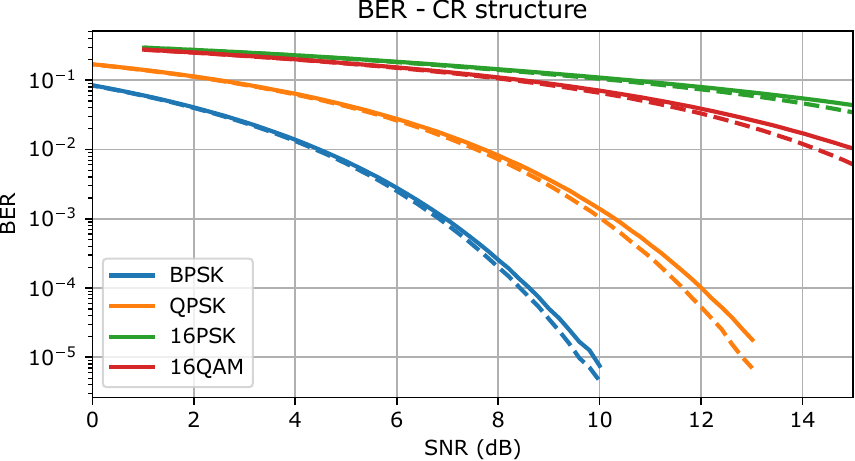}
    \caption{BER against SNR with BPSK, QPSK, 16-PSK and 16-QAM modulation, with the CR structure (solid lines) and without any interference (dashed lines).}
    \label{fig:RIC_BER}
\end{figure}

In order to evaluate the achieved performance, a scenario is defined in Table \ref{tab:scenar1}. Fig. \ref{fig:RIC_P_D} illustrates the detection probability obtained for different modulation indices, i.e. BPSK, QPSK, 16-PSK and 16-QAM, after interference cancellation at the radar function, compared to the detection probability obtained when there is no interference, or when interference is not cancelled. In the considered scenario, the power of the communication signal is 20 dB higher than radar echoes. Even if the interference power is spread in the delay-Doppler map owing to the presence of uncompensated communication symbols, the radar echoes are hidden below the interference noise floor. Thanks to interference cancellation, large performance improvement is achieved, depending on the BER obtained at the communication function (and equivalently on the  modulation indices), as shown in Fig. \ref{fig:RIC_BER}. At low communication SNR, the BER is high, leading to multiple symbol errors at the cancellation. The higher the modulation index, the higher the number of errors, leading to a decrease of the detection probability. In that case, the obtained performance may even be worse than the performance obtained when there is no interference cancellation with high modulation indices. However, at higher communication SNR, the BER decreases, and the cancellation efficiency increases as well. Therefore, the detection probability increases to one. The increase is slower for high modulation indices since the decrease BER is slower in the considered range of SNRs. Regarding the 16-QAM constellation, the detection performance is worse than the one achieved with the 16-PSK for SNRs up to around 13 dB, even with a lower BER, owing to the non-constant envelope of the transmitted ISAC waveform.

\subsection{Introduction of Error Correcting Codes}
In order to further improve the interference cancellation capabilities, Error Correcting Codes (ECC), e.g. convolutional codes, may be introduced in the structure. In this paper, bits are coded at the transmitter using the systematic convolutional code $\textbf{G}(D) = \left[1 \quad \frac{1+D}{1+D+D^2}\right]$ of rate $R_c = 1/2$, with $D$ the delay operator. At the receiver, soft Viterbi decoding is performed.\smallskip

First, regarding the inherent interference mitigation of the PC-FMCW ISAC waveform discussed in Section \ref{sec:RADAR_int_theoretical_analysis}, the auto-correlation of $z_\C$ presented in \eqref{eq:interference_radar_autocorr} remains valid if the compensated symbols remain independent with each other, such that \eqref{eq:cond_symbols} is still fulfilled. Apart from repetition coding, even if convolutional codes integrate redundancy in the coded bits, pairs of symbols remain empirically independent, except at the start of the sequence. Interleaving can also be introduced to ensure that behaviour.  Therefore, with the considered convolutional code, the dispersion of the interfering communication signal in the Delay-Doppler remains unchanged.\smallskip

\begin{figure}
\centering
\includegraphics[width=\linewidth]{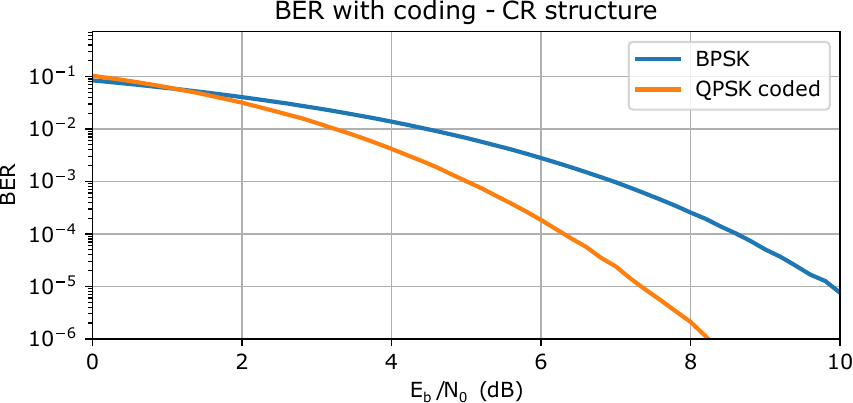}
\caption{BER against $\EbNo$ for uncoded BPSK and coded QPSK communication symbols.}
\label{fig:RIC_BER_coded}
\vspace*{0.2cm}

\includegraphics[width=\linewidth]{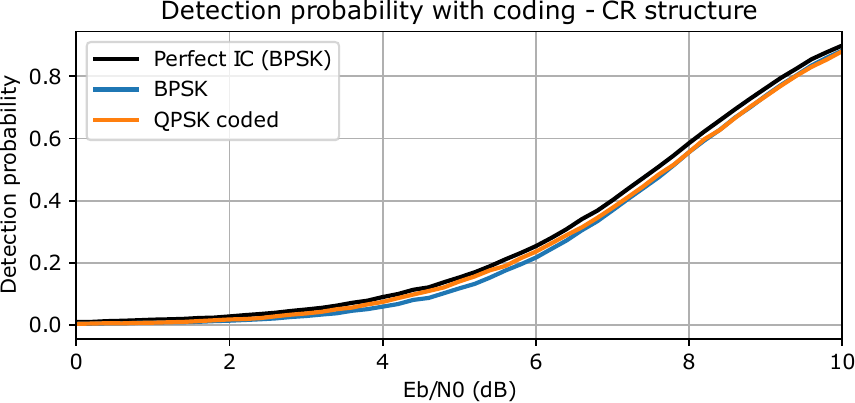}
\vspace*{-0.15cm}

\includegraphics[width=\linewidth]{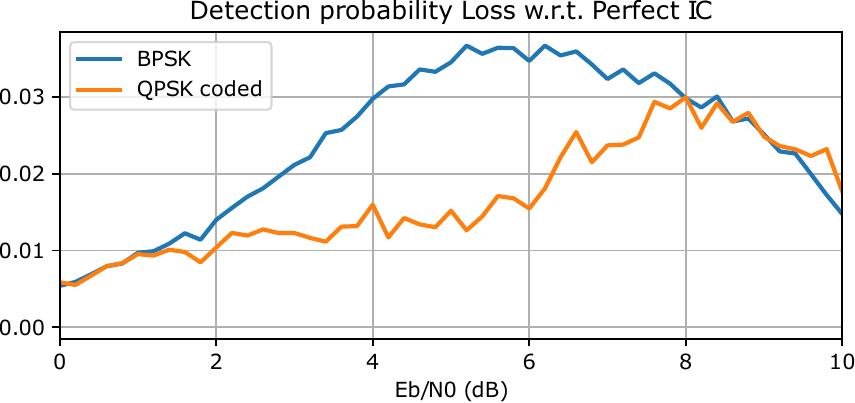}
\caption{Detection probability against $\EbNo$ after interference cancellation, for uncoded BPSK and coded QPSK communication symbols.}
\label{fig:RIC_P_D_coded}
\end{figure}

In order to ensure that a fair comparison is performed between the uncoded and coded transmissions, coded bits are transmitted with half power in the coded transmission in order to keep a constant energy per bit $E_\text{b}$. The uncoded transmission uses BPSK constellation, whereas the coded transmission uses QPSK constellation in which the redundancy bits are transmitted in quadrature, ensuring the same transmission time. This also enables to conserve the same $\text{SNR} = M R_c \EbNo$ in both transmissions, with $M$ the modulation index ($M=1$ for BPSK, and $M=2$ for QPSK)\smallskip

Fig. \ref{fig:RIC_BER_coded} and \ref{fig:RIC_P_D_coded} respectively illustrate the achieved BER and detection probability with uncoded BPSK and coded QPSK transmissions. On the one hand, at the communication function, a lower BER is achieved with the coded transmission thanks to the presence of ECC in the ISAC waveform, except for low $\EbNo$ ratios. On the other hand, for the radar function, the detection probability achieved with the uncoded BPSK is close to the performance achieved without any interference in the considered scenario. In fact, the BER of the uncoded BPSK transmission is small w.r.t. the number of transmitted symbols per pulse for $\EbNo$ ratio higher than 4 dB. Yet, a slightly higher detection probability is achieved with the coded QPSK transmission in that range of $\EbNo$ ratio, owing to the BER improvement with ECCs. Note that the same could be achieved with higher order constellations, if codes with higher correction capabilities are considered.

\subsection{Summary}
When the communication component is dominant compared to the radar echoes, interference cancellation performs well when the BER obtained at the communication output is sufficiently low. The modulation index of the communication signal can be increased at the price of a lower detection probability for a given SNR at the communication function. Introducing ECCs helps to improve the achieved performance at both functions.

\section{Interference Cancellation at the Communication Function}
\label{sec:interference_comm}
Let us now focus on the communication function. The scenario presented in Fig. \ref{fig:FD_scenario} is still considered, in which target echoes are received simultaneously with a communication signal coming from another system sharing the same system parameters. At the communication function, the signal of interest is the communication signal, and the radar echoes reflected by the targets are interfering. 

\subsection{Theoretical Analysis}
As for the radar receiver, dechirping is applied at the communication receiver. Hence, the interfering components rotate owing to the Doppler frequencies of the paths, in addition to the generated beat frequencies. However, since the ISAC systems parameters are considered to be identical for every system, there is no inherent interference mitigation in the processing of the communication function. Therefore, interference cancellation must be applied to increase the communication performance if interference occurs.

\begin{table}
    \centering
    \caption{Simulation parameters for Section \ref{sec:interference_comm}. The false alarm probability requirement is denoted by $\mathcal{P}_\text{FA}$.}
    \renewcommand{\arraystretch}{1.0}
    \begin{tabular}{|l|l|l|}
        \hline
        \multirow{2}{*}{$B$ = 20 MHz} &\multirow{ 2}{*}{$T$ = 100 $\mu$s} & I: $T_g$ = 0.5 $\mu$s \\
         && II: $T_g$ = 5 $\mu$s \\\hline
         \multicolumn{1}{|c|}{$L_c$ = 100} & \multicolumn{1}{c|}{$\mathcal{P}_\text{FA} = 10^{-4}$} & \multicolumn{1}{c|}{QPSK modulation}  \\\cline{1-3}
        \multirow{2}{*}{$|\alpha_{\R}|$ = 1} & I: $\tau_{\R}$ = 0.1 $\mu$s & \multirow{2}{*}{$f_{\mathrm{D}\R}$ = 1000 Hz} \\
        & II: $\tau_{\R}$ = 1 $\mu$s & \\\hline
    \multirow{2}{*}{$|\alpha_\C|$ = 1} & I: $\tau_\C$ = 0.25 $\mu$s & \multirow{2}{*}{$f_{\D\C}$ = -300 Hz} \\
        & II: $\tau_\C$ = 2.5 $\mu$s & \\\hline
    \end{tabular}
    \renewcommand{\arraystretch}{1.}
    \label{tab:scenar2}
    \end{table}
    
    \begin{figure}
    \centering
    \includegraphics[width=\linewidth]{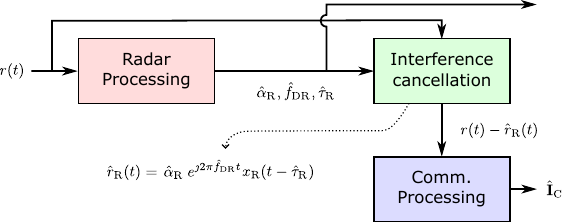}
    \caption{Receiver chain implementing the RC structure. The radar function forwards the transmitted symbols, and  the estimates of the attenuation, delay and Doppler frequency of the radar echo to the communication function.}
    \label{fig:Comm_IC_chain}
\end{figure}

\subsection{Interference Cancellation}
In scenarios where the power of the radar echo is assumed to be larger than the power of the communication signal, interference cancellation can be applied at the communication function. Following the processing of Section \ref{sec:radar_processing}, from the received signal $r$, the radar echo parameters, i.e. $\alpha_\R$, $f_{\text{DR}}$ and $\tau_\R$, are first estimated. Then, In addition to the transmitted symbols $\textbf{I}_\R$, all these parameters are forwarded to the interference cancellation block of the communication function, which generated an estimate of the signal $r_\R$:
\begin{equation}
\hat{r}_\R(t) = \hat{\alpha}_\R \: e^{j 2\pi \hat{f}_\text{DR} t} x_\R(t-\hat{\tau}_\R),
\end{equation}
where the symbol $\hat{\cdot}$ designates estimated quantities. Then the communication processing is applied on the received signal $r$, from which the radar echo has been suppressed: 
\begin{equation}
	\tilde{r}_\C(t) = r(t) - \hat{r}_\R(t).
\end{equation}
This receiver is named \textit{RC structure}, also based on the order in which the radar and communication functions are processed. It is illustrated in Fig. \ref{fig:Comm_IC_chain}. Note that this process can be repeated for each radar echo if there is multiple targets. In that case, the complex coefficients $\alpha_{\R k}$ can be estimated from the delay-Doppler map, but this may be inaccurate, owing to the inter-target interference. Instead, thanks to the joint waveform structure which is shared by every system, radar echoes can be processed as communication signals, for which the delays and the Doppler frequencies have already been estimated at the radar function. Additionally, the transmitted symbols are also known. Another option is to estimate all the parameters of each radar echo successively based on the estimated number of targets of the radar receiver, but this solution does not leverage on all the information gathered at the radar function.
\smallskip

When the estimations of the delay and Doppler frequency are based on the radar receiver output, the achieved performance depends on the accuracy of the radar estimates, which is directly related to the delay and Doppler resolutions of the radar function. In order to increase the Doppler resolution, one can increase the total duration of the payload. Nevertheless, this also impacts the radar processing gain, and the sensitivity to the frequency offset. Otherwise, the accuracy can be improved by interpolating the delay-Doppler map through zero-padding. Two scenarios have been defined in Table \ref{tab:scenar2} to evaluate the impact of the Doppler resolution on the achieved performance.\smallskip

\begin{table}
\centering 
\caption{Resolution and accuracy achieved for both scenarios of Table \ref{tab:scenar2}.}
\begin{tabular}{|c|c|c|c|c|}
\hline
\multirow{2}{*}{\makecell{\textbf{Number of} \\ \textbf{pulses} $P$}} & \multicolumn{2}{c|}{\textbf{Resolution} ($1/T_\text{CPI}$)} & \multicolumn{2}{c|}{\textbf{Accuracy}} \\\cline{2-5}
& Scenario I & Scenario II & Scenario I & Scenario II \\\hline\hline
50 & 199 Hz & 190.5 Hz & 5 Hz & 47.6 Hz \\
100 & 99.5 Hz & 85.2 Hz & 5 Hz & 47.6 Hz \\
150 & 66.3 Hz & 63.5 Hz & 5 Hz & 15.9 Hz \\
200 & 49.8 Hz & 47.6 Hz & 5 Hz & 0 Hz \\\hline
\end{tabular}
\label{tab:resolution}
\end{table}

\begin{figure}
    \centering
    \includegraphics[width=\linewidth]{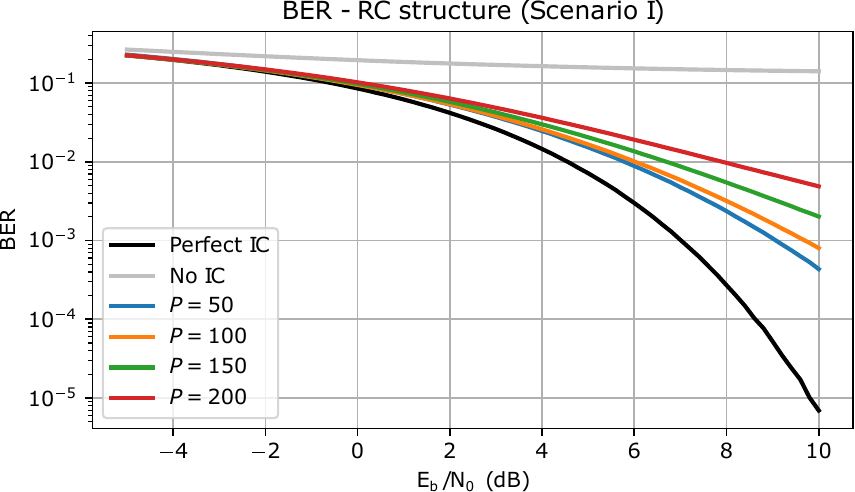}
    \caption{BER against $\EbNo$ for multiple pulse numbers.}
    \label{fig:CIC_BER_sc1}
    \vspace*{0.4cm}
    
    \includegraphics[width=\linewidth]{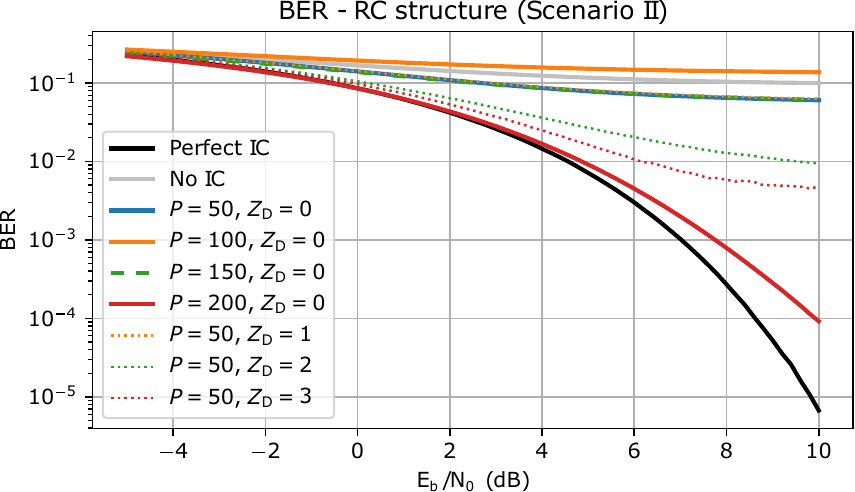}
    \caption{BER against $\EbNo$ for multiple pulse numbers and zero-padding factors ($Z_\D$ denotes the number of padded zeros).}
    \label{fig:CIC_BER_sc2}
    \vspace*{0.4cm}
    
    \includegraphics[width=\linewidth]{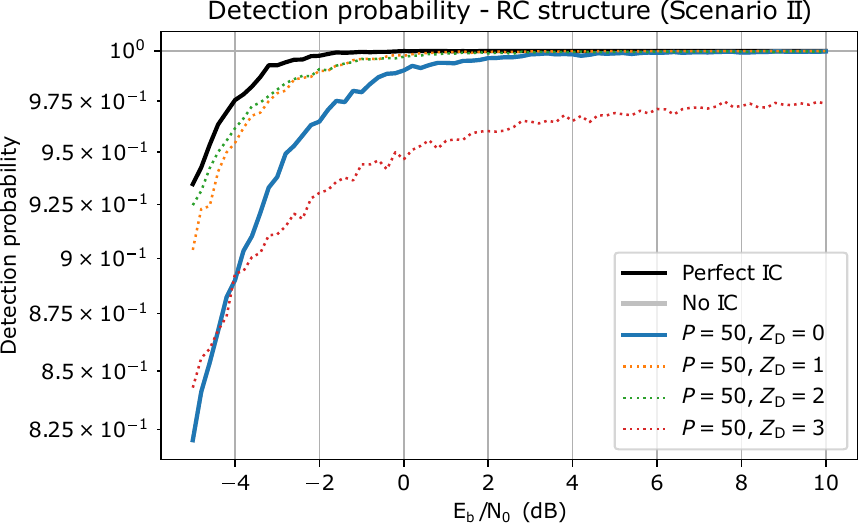}
    \caption{Detection probability against $\EbNo$ for multiple zero-padding factors ($Z_\D$ denotes the number of padded zeros).}
    \label{fig:CIC_P_D_sc2}
\end{figure}

The higher the number of transmitted pulses $P$, the better the Doppler resolution. Yet, this does not imply a higher accuracy. The accuracy depends on the Doppler frequency of the target, w.r.t. the discrete delay-Doppler grid. For instance, Table \ref{tab:resolution} summarises the achieved resolutions and accuracies for both scenarios defined in Table \ref{tab:scenar2}. On the one hand, with the parameters of the first scenario, even if the resolution is improving with $P$, the accuracy remains constant. Yet, the length of the payload also increases, leading to many errors at the reconstruction of the end of the signal owing to the phase rotation of the symbols. This is translated in Fig. \ref{fig:CIC_BER_sc1}, which shows the BER achieved with the RC structure in the first scenario, for different pulse numbers. The detection probability is not shown as it is always very close to one in the considered scenario. Increasing the number of transmitted pulses leads to an increase of the BER.\smallskip

On the other hand, with the parameters of the second scenario, the accuracy remains constant when 50 or 100 pulses are transmitted, but decreases for higher number of pulses. Fig. \ref{fig:CIC_BER_sc2} illustrates the BER achieved with the RC structure in the second scenario, for different pulse numbers and zero-padding factors. Consequently,
\begin{itemize}
\item when the number of pulses is increased from 50 to 100 pulses, worse performance is achieved;
\item when the number of pulses is increased from 100 to 150 pulses, the accuracy improvement enhances the BER. Yet, the increase of payload duration worsens the performance, leading to a limited improvement w.r.t. the transmission of 100 pulses; 
\item when the number of pulses is increased from 150 to 200 pulses, the Doppler frequency is estimated correctly, leading to a performance improvement, regardless of the increased payload duration.
\end{itemize} 
\smallskip

With zero-padding, the accuracy is increased without affecting the payload length. For instance, with 100 transmitted pulses, a zero-padding factor of 1 does not increase the accuracy, but the payload length being unaffected,  performance losses are avoided. With higher zero-padding factors, the accuracy is increased, and better performance is achieved. Note that a saturation of the detection probability is observed with a zero-padding factor of 3, which is translated by a saturation of the BER, and lower performance compared to the transmission of 200 pulses. This is a consequence of the CFAR detector, which is not able to estimate accurately the noise power in the delay-Doppler map with high zero-padding factors, owing to the interpolation of the noise and target response.

\subsection{Summary}
When the radar echoes are dominant compared to the communication signal, interference cancellation can be applied efficiently to the communication receiver to cancel the radar echoes. The estimation of the echoes parameters is either based on the radar outputs, or on the communication synchronisation algorithm. In the former, the radar delay and Doppler resolutions should be sufficiently high in order to ensure accurate estimations of the delays and Doppler frequencies, leading to an efficient cancellation. The resolutions can be increased by modifying the system parameters. Otherwise, zero-padding of the delay-Doppler map is helpful to increase the accuracy of the delay and Doppler estimations without modifying the system parameters.

\section{Iterative Interference Cancellation}
\label{sec:iterative_IC}
In Sections \ref{sec:interference_RADAR} and \ref{sec:interference_comm}, two interference cancellation structures at the radar or communication function have been detailed, namely the CR and RC structures. However, each receiver only works well respectively at high communication or radar Signal to Interference Ratio (SIR). Thus, interference cancellation should be integrated at the radar (resp. communication) function only if the communication signal (resp. radar echo) is dominant compared to the radar echo (resp. communication signal). To solve this issue, if the power levels of the different signals are estimated beforehand, one can switch of ISAC structure to use the appropriate interference cancellation scheme. Another solution which does not require any a priori information is to perform iterative cancellation.\smallskip

\begin{figure}
    \centering
    \includegraphics[width=\linewidth]{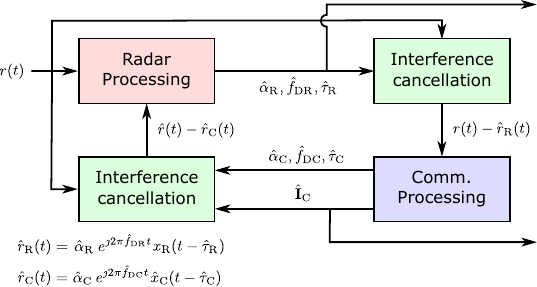}
    \caption{Receiver chain implementing the R-iterative structure. The C-iterative structure is implemented similarly, except that the communication processing is performed first.}
    \label{fig:iterative_structure}
\end{figure}

Let us respectively denote the radar and communication functions by R and C. The extension of the RC and CR receiver into iterative structures is illustrated in Fig. \ref{fig:paper_structure}. In iterative structures, the received signal is forwarded to the first radar or communication block, and to interference cancellation blocks. Then, multiple iterations of both functions are performed, with interference cancellation blocks inserted before each function. We respectively name these receivers \textit{R- and C-iterative structures}, when they start with a radar or communication block, respectively. The R-iterative structure is illustrated in Figure \ref{fig:iterative_structure}.\smallskip 

In this paper, multiple specific iterative interference cancellation receivers are defined, namely the RCR (extending the CR structure with one additional radar function), CRC (extending the RC structure with one additional communication function), and RCRC (R-iterative receiver with two complete stages) structures. Note that there is no benefit to perform more than two complete stages since there are no new operations performed w.r.t. the two-stage iterative structure. Parallel interference cancellation receivers can also be defined by combining simple or iterative structures. Nonetheless, all the proposed structures require multiple radar and communication processing, coupled with multiple interference cancellation stages. Thus, the complexity of these receivers is higher than the simple structures presented in Section \ref{sec:interference_RADAR} and \ref{sec:interference_comm}.

\subsection{Numerical Analysis}

\begin{table}[ht]
    \centering
    \caption{Simulation parameters for Section \ref{sec:iterative_IC}. The false alarm probability requirement is denoted by $\mathcal{P}_\text{FA}$.}
    \begin{tabular}{|l|l|l|}
    \hline
    $B$ = 20 MHz & $T$ = 100 $\mu$s & $T_g$ = 5 $\mu$s \\\hline
    $P$ = 50 & $L_c$ = 100 & $\EbNo$ = 10 dB \\\hline
    $\mathcal{P}_\text{FA} = 10^{-4}$ & QPSK modulation & $Z_\D$ = 3 \\\hline
    $|\alpha_{\R}|$ = 0.1$\rightarrow$10 & $\tau_{\R}$ =1 $\mu$s & $f_{\mathrm{D}\R}$ = 1000 Hz \\\hline
    $|\alpha_\C|$ = 1 & $\tau_\C$ = 2.5 $\mu$s & $f_{\D\C}$ = -300 Hz \\\hline
    \end{tabular}
    \label{tab:scenarRCIC}
\end{table}

\begin{figure}
    \centering
    \includegraphics[width=\linewidth]{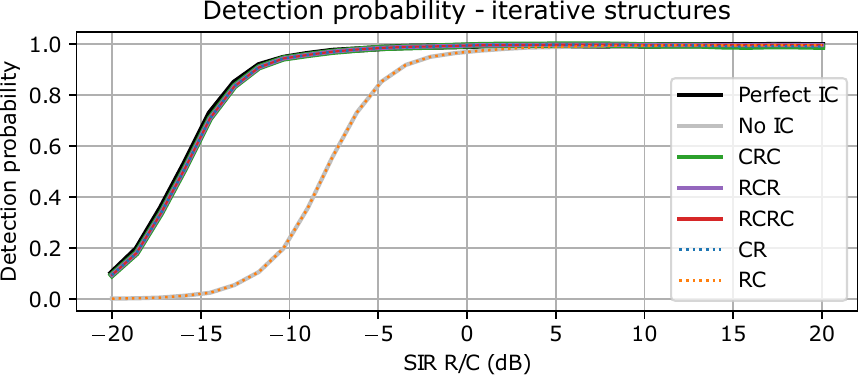}
    \caption{Detection probability against radar echo to communication signal power ratio, with simple and iterative structures.}
    \label{fig:iterative_P_D}
    \vspace*{0.4cm}

    \includegraphics[width=\linewidth]{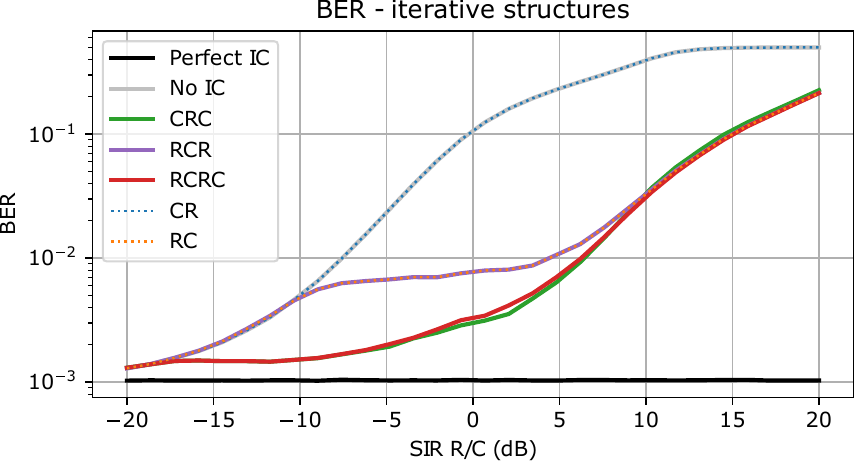}
    \caption{BER against radar echo to communication signal power ratio, with simple and iterative structures.}
    \label{fig:iterative_BER}
\end{figure}

Figures \ref{fig:iterative_P_D} and \ref{fig:iterative_BER} illustrate the achieved detection probability and BER against different SIRs, i.e. radar echo to communication signal power ratios, for a scenario depicted in Table \ref{tab:scenarRCIC}. Three regions can be identified.

\paragraph{For radar SIRs between -20 dB and -10 dB} The communication signal is dominant compared to the radar echo. Therefore, low detection probabilities are achieved without interference cancellation before the radar function, as with the RC structure. The other simple and iterative structures nonetheless achieve good detection probabilities since there is at least one interference cancellation stage before the radar function. Regarding the BER, the CR, RC and RCR structures achieve similar performance, with a slight increase of BER for the RC structure, owing to the poor radar and interference cancellation performed at the first stage. However, the CRC and RCRC structures achieve better performance with an increasing power of the radar echo, since the radar function is improved by the first communication and interference cancellation stage, and then the communication performance are improved at the second stage by the radar function and interference cancellation block.

\paragraph{For radar SIRs between -10 dB and 0 dB} The radar echo power increases to reach power levels equivalent to the communication signal. Owing to the high radar processing gain, the RC and all the iterative structures perform well compared the CR structure, even if the SIR is still negative. The BERs achieved by the CRC and RCRC structures starts to increase with the power of the radar echo since the interference cancellation is not perfect, and the residual radar interference is increasing with this power. Still, the RCR structure achieves lower performance compared to the other iterative structures because of a poorer radar detection and interference cancellation at the first stage. This first stage is nonetheless improving with the power of the radar echo and performance does not decrease as fast as the other structures.

\paragraph{For radar SIRs between 0 dB and 20 dB} The power of the radar echo is dominant compared to the communication signal. All the structures efficiently suppress the radar interference at the communication function, but the residual interference continues to increase with the radar echo power, and the achieved BERs thus increase. Furthermore, at SIRs close to 20 dB, the detection probabilities of the CR and CRC structures decreases slightly owing to the poor communication and interference cancellation performed at the first stage.\\

Finally, note that the RCRC structure achieves slightly lower performance compared to the CRC structure between -5 dB and 5 dB, owing to the additional interference cancellation residual from the first stage, which propagates through the complete structure.

\subsection{Summary}
Compared to simple structures, iterative structures enable to achieve good performance at the radar and communication function, irrespective of the considered scenario. Multiple iterative structures have been defined with different complexity and performance. The CRC receiver seems to achieve the best radar and communication performance. Still, interference cancellation is not perfect, and the residual interference of the radar echo on the communication signal, increasing with the power of the radar echo, is detrimental for the communication function at high SIRs.

\section{Interference Cancellation in Dynamic Scenarios}
\label{sec:dynamic_IC}
Compared to the simple structures, the iterative interference cancellation structures presented in Section \ref{sec:iterative_IC} are efficient in any scenario, and does not require any prior estimation of the signals power levels. Yet, the introduction of multiple interference cancellation stages increases their complexity. In dynamic scenarios, one can leverage on previous radar measurements to estimate the power of the signals, or to reduce the iterative structures complexity. This is done for instance in \cite{09557830}, which predicts the vehicles states based on their estimated kinematic parameters, and infers delays and Doppler frequencies associated to different channel paths. In this section, a focus is made on the R-iterative structures, for which the first radar function is replaced by an estimation of the radar echo parameters, computed as an update of previous radar processings. These structures are respectively named \textit{dynamic CR} and \textit{dynamic CRC receivers}.

\subsection{System Model}

\begin{figure}
\centering
\includegraphics[width=\linewidth]{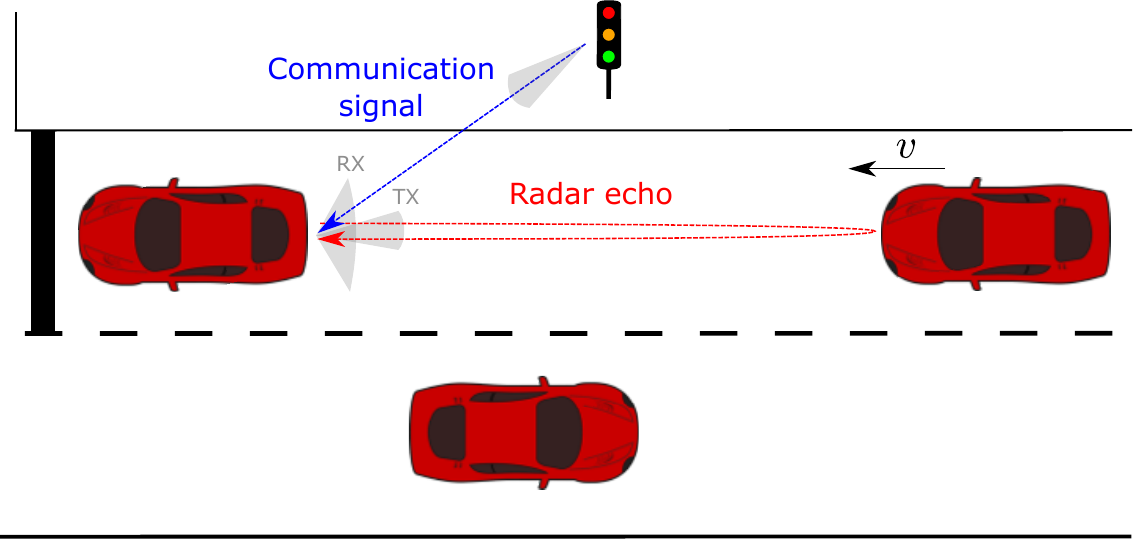}
\caption{Considered vehicular scenario. A vehicle is receiving a communication signal from a RoadSide Unit (RSU), while detecting the next vehicle behind, which is braking.}
\label{fig:scenario_vehicular}
\end{figure}

The vehicular scenario of Fig. \ref{fig:scenario_vehicular} is considered. The vehicle of interest, located in $(0,0)$, is stopped on the road. It receives simultaneously a communication signal from a RSU, and a radar echo from the vehicle behind, which is braking. Denoting by $\textbf{p}_\C$ and $\textbf{p}_\R(n)$ the 2D positions of the RSU and the moving vehicle at the $n^\text{th}$ transmission, the links parameters are computed as 
\begin{align}
\nonumber &\tau_\C = \frac{\rVert \textbf{p}_\C\rVert}{c},\: \tau_\R(n) = \frac{2\rVert \textbf{p}_\R(n)\rVert}{c},\: f_{\D\R}(n) = \frac{2 f_c v(n)}{c}, \\
& \alpha_\C(n) = \frac{c \: \exp{-j2\pi f_c \tau_\C}}{4\pi f_c \rVert \textbf{p}_\C \rVert} , \: \alpha_\R(n) = \frac{c \rho \: \exp{-j2\pi f_c \tau_\R(n)}}{4\pi f_c (2\rVert \textbf{p}_\R(n) \rVert)}, \label{eq:alpha_vehicle}
\end{align}
where $v(n)$ is the speed of the moving vehicle at time index $n$, $f_c$ is the carrier frequency and $c$ the speed of light. The considered radar propagation model is a geometrical optics-based model, with $\rho$ being the Fresnel reflection coefficient of the target. Assuming a good conductor, it is approximated as $\rho \approx 1$. This model is suited for automotive radars owing to the high carrier frequencies, meaning that near-field propagation should be considered. Note that the same analysis can be performed with other propagation models, and fading on the links can also be introduced without discarding the following discussions. With the attenuations of \eqref{eq:alpha_vehicle}, the radar SIR is computed as $\text{SIR}(n) = \rVert \textbf{p}_\C\rVert^2 / 4\rVert \textbf{p}_\R(n) \rVert^2$. Let us assume a time interval $\Delta_t$ between two radar measurements. In the considered scenario, the estimates of the delays and Doppler frequencies at the next transmission are computed as
\begin{equation}
    \left\{
\begin{alignedat}{10}
&\hat{f}_{\D\R}(n+1) = \hat{f}_{\D\R}(n), \\
&\hat{\tau}_{\R}(n+1) = \hat{\tau}_{\R}(n) + \frac{\Delta_t}{2c}\: \hat{v}(n) = \hat{\tau}_{\R}(n) + \frac{\Delta_t}{f_c} \: \hat{f}_{\D\R}(n).
\end{alignedat}\right.
\end{equation}
This simple estimation update process could be improved, for instance using complex tracking mechanisms, as extended Kalman filtering \cite{09171304}. Nonetheless, the proposed process is sufficient to achieve good performance.

\subsection{Numerical Analysis}

\begin{table}
    \centering
    \caption{Simulation parameters for Section \ref{sec:dynamic_IC}. The false alarm probability requirement is denoted by $\mathcal{P}_\text{FA}$.}
    \begin{tabular}{|l|l|l|}
    \hline
    $B$ = 20 MHz & $T$ = 100 $\mu$s & $T_g$ = 5 $\mu$s \\\hline
    $P$ = 50 & $L_c$ = 100 & $\EbNo$ = 10 dB \\\hline
    $\mathcal{P}_\text{FA} = 10^{-4}$ & QPSK modulation & $Z_\D$ = 3 \\\hline
    $\textbf{p}_{\R}(0) = (30,0)$ m & $\textbf{p}_{\C} =  (14,2.5)$ m & $v(n)=v=15$ m/s \\\hline
    \end{tabular}
    \label{tab:scenar_dyn_IC}
    \end{table}
    
\begin{figure}
    \centering
    \includegraphics[width=\linewidth]{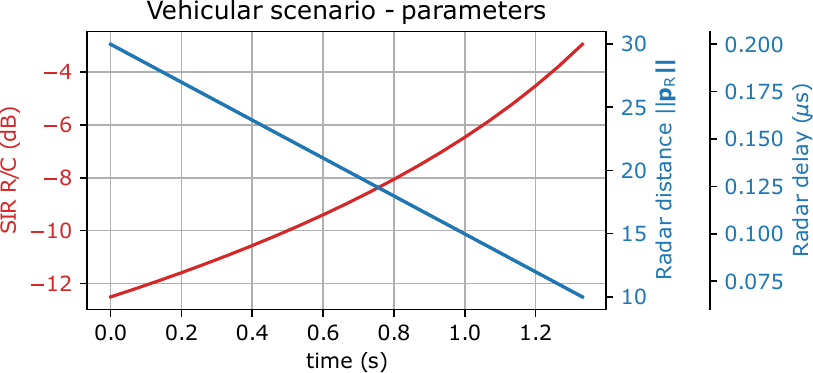}
    \caption{Vehicular scenario parameters against time.}
    \label{fig:dynamic_parameters}
\end{figure}

\begin{figure}
    \centering
    \includegraphics[width=\linewidth]{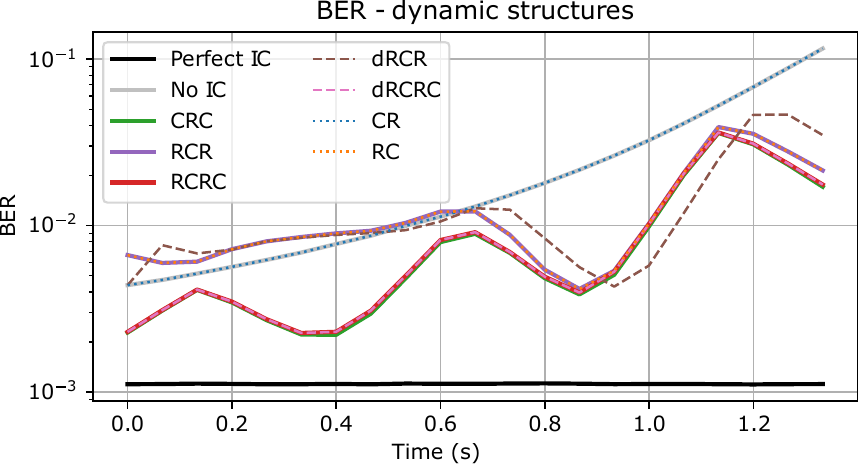}
    \caption{BER against time, with iterative and dynamic structures.}
    \label{fig:dynamic_BER}
\end{figure}

\begin{figure}
    \centering
    \includegraphics[width=\linewidth]{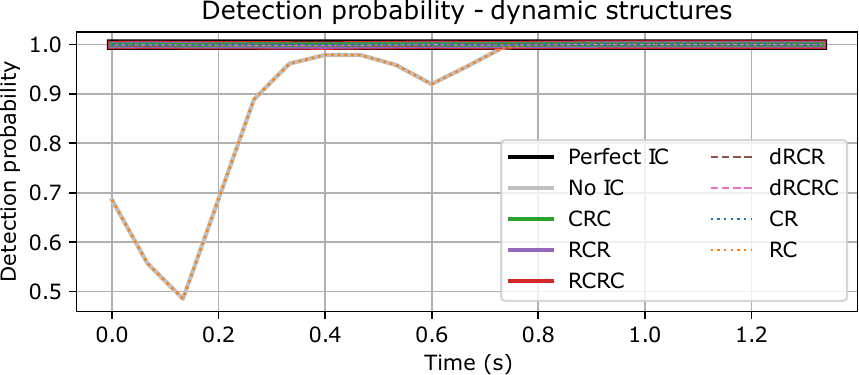}
    \caption{Detection probability against time, with iterative and dynamic structures.}
    \label{fig:dynamic_P_D}
\end{figure}

Table \ref{tab:scenar_dyn_IC} depicts the scenario parameters, and their evolutions is shown in Fig. \ref{fig:dynamic_parameters}. The delay between two transmissions is set as $\Delta_t = 66.6$ ms. At the start, the communication signal power dominates. As time goes by, the moving vehicle approaches the motionless vehicle, leading to an increased radar echo power $|\alpha_\R(n)|^2$, and a decreased delay $\tau_\R(n)$. Fig. \ref{fig:dynamic_BER} and \ref{fig:dynamic_P_D} illustrate the achieved detection probability and BER against time. First, regarding the detection probability, oscillations are observed with time owing to the finite radar delay accuracy. Except for the RC structure, all other simple, iterative, and dynamic receivers achieve a detection probability close to one, thanks to the interference cancellation of the communication signal. Next, regarding the BER, two cases can be distinguished:
\begin{itemize}
\item Until $t\approx 0.7$ s, the RC, RCR, and dynamic RCR structures achieve a slightly higher BER than the other structures, which is also higher than the one achieved with the structure without any interference cancellation of the radar echo. At low radar echo to communication signal power ratio, these structures are hugely impacted by the communication signal when performing the first radar and interference cancellation stage. The reconstruction and cancellation of the radar echo is thus inefficient, leading to an increased interference level at the communication function.
\item For higher times, the performance of all structures are similar or better than the performance of the CR structure. The radar echo power is still lower than the communication signal, but it becomes sufficient to perform an efficient radar and interference cancellation stage, even without suppressing the communication signal beforehand. Thus, at such times, starting by the radar function is no more detrimental.
\end{itemize}
Regarding the dynamic structures, on the one hand, the dynamic RCRC receiver achieves the same radar and communication performance as the RCRC structure. The first radar and interference cancellation stage of this structure has nearly no impact on the last communication block, as shown by the CRC structure which achieves the same performance. On the other hand, this is not the case with the dynamic CRC structure. The performance observed with this structure are shifted in time compared to the performance achieved by the CRC structure. This is a consequence of the update step, which relates the efficiency of the previous to the actual time instant. When the radar function performs well (resp. poorly), the estimation of the delay and Doppler frequency at the next time instant is accurate (resp. inaccurate), leading to good (resp. low) performance for the next transmission.

\subsection{Summary}
Leveraging on multiple successive transmissions in dynamic scenarios, the complexity of R-iterative structures is reduced by replacing the first radar block by an update of previous radar estimations, at the price of a slight performance reduction. One can still improve the performance by using more complex tracking mechanisms, or by enhancing the radar resolution. 

\section{Conclusion}
\label{section:conclusion}
Interference between multiple DFRC systems in full-duplex is a major issue for the future, especially for automotive scenarios. Whereas an inherent interference mitigation is present at the radar function thanks to the introduction of communication symbols, there is no inherent mitigation at the communication function. In order to improve both radar and communication performance, interference mitigation has been introduced. First, simple structures in which interference cancellation is only applied at the communication or radar function have been presented. With such structures, good performance is achieved, respectively in scenarios where the radar echoes or the communication signal are dominant. The impact of the system parameters and error correcting codes on the performance has also been analysed. Then, iterative interference cancellation structures have been introduced. It has been shown that it increases the robustness of the system to any scenario. Finally, a dynamic automotive scenario has been considered, in which previous radar measurements are used in order to reduce the complexity of the R-iterative structures, at the price of a slight performance reduction.\smallskip

In future works, the proposed interference cancellation schemes could be extended for Multiple Input Multiple Output (MIMO) ISAC systems. For instance, this would enable a comparison between interference cancellation and beamforming techniques for interference mitigation in full-duplex ISAC scenarios. Other performance metrics could also be analysed to obtain a complete analysis of the comparison between full-duplex systems with interference cancellation or/and beamforming, and half-duplex systems implementing time or frequency division duplexing.

\bibliographystyle{ieeetr}
\bibliography{biblio}

\begin{IEEEbiography}[{\includegraphics[width=1in,height=1.25in,clip,keepaspectratio]{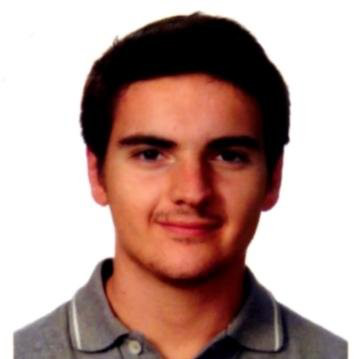}}]{François De Saint Moulin} received the B.Sc. and the M.Sc. degrees in 2018 and 2020, respectively, in electrical engineering from the Université catholique de Louvain, (UCLouvain), Louvain-la-Neuve, Belgium, where he is currently working toward the Ph.D. degree with the Institute for Information and Communication Technologies, Electronics and Applied Mathematics, (ICTEAM), UCLouvain. His research interests include integrated sensing and communication technologies, near-field channel modelling and signal processing for radar and communication, and stochastic geometry.
\end{IEEEbiography}

\begin{IEEEbiography}[{\includegraphics[width=1in,height=1.25in,clip,keepaspectratio]{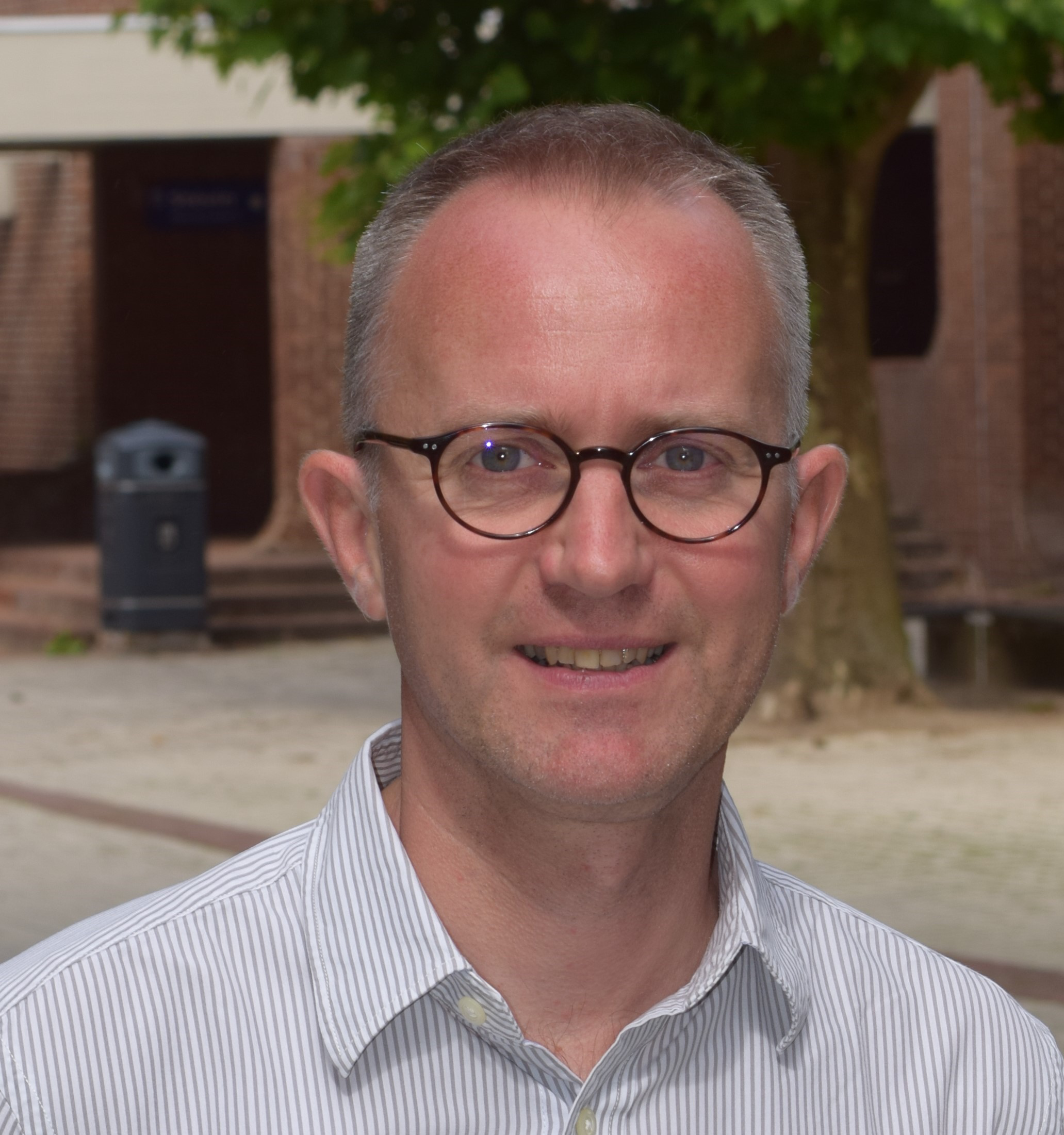}}]{Claude Oestges} (Fellow, IEEE) received the M.Sc. and Ph.D. degrees in Electrical Engineering from UCLouvain, Louvain-la-Neuve, Belgium, respectively in 1996 and 2000. In January 2001, he joined as a post-doctoral scholar the Smart Antennas Research Group (Information Systems Laboratory), Stanford University, CA, USA. From January 2002 to September 2005, he was associated with the Microwave Laboratory UCLouvain as a post-doctoral fellow of the Belgian Fonds de la Recherche Scientifique (FRS-FNRS). Claude Oestges  is presently Full Professor with  the Electrical Engineering Department, Institute for Information and Communication Technologies, Electronics and Applied Mathematics (ICTEAM), UCLouvain. He is the author or co-author of four books and more than 200 journal papers and conference communications, and he was the chair of COST Action CA15104 IRACON (2016-2020), as well as the recipient of the 1999-2000 IET Marconi Premium Award and of the IEEE Vehicular Technology Society Neal Shepherd Award in 2004 and 2012.
\end{IEEEbiography}

\begin{IEEEbiography}[{\includegraphics[width=1in,height=1.25in,clip,keepaspectratio]{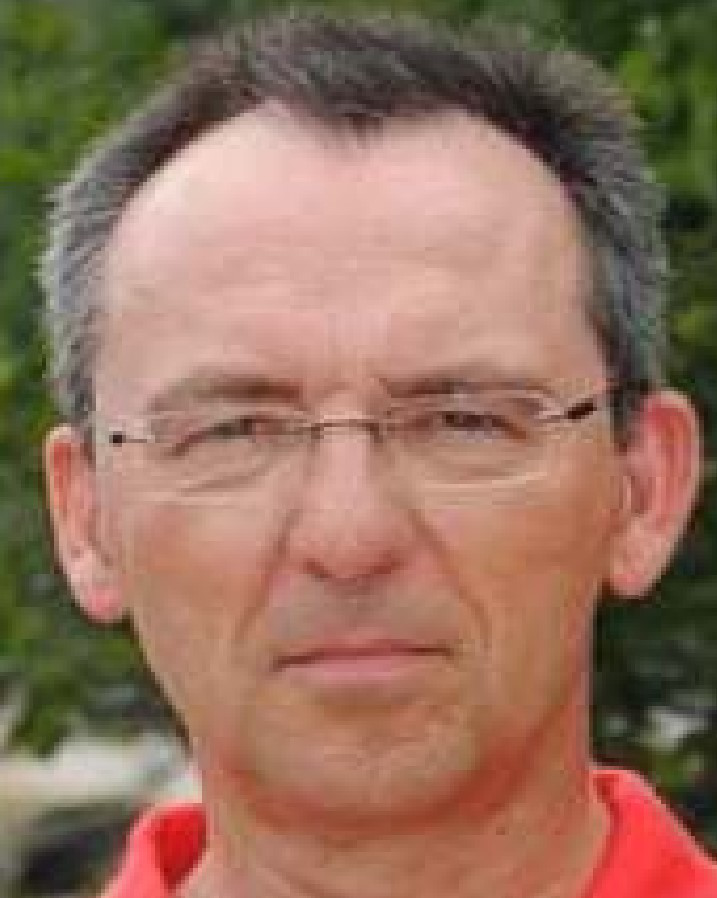}}]{Luc Vandendorpe} (Fellow, IEEE) was born in Mouscron, Belgium, in 1962. He received the degree (summa cum laude) in electrical engineering and the Ph.D. degree in applied science from UCLouvain, Louvain-la-Neuve, Belgium, in 1985 and 1991, respectively. Since 1985, he has been with the Communications and Remote Sensing Laboratory, UCL, where he first worked in the field of bit rate reduction techniques for video coding. In 1992, he was a Visiting Scientist and a Research Fellow with the Telecommunications and Traffic Control Systems Group, Delft Technical University, The Netherlands, where he worked on spread spectrum techniques for personal communications systems. From October 1992 to August 1997, he was a Senior Research Associate with the Belgian NSF, UCL, and an invited Assistant Professor. He is currently a Full Professor with the Institute for Information and Communication Technologies, Electronics and Applied Mathematics, UCLouvain. His research interests include digital communication systems and more precisely resource allocation for OFDM(A)-based multicell systems, MIMO and distributed MIMO, sensor networks, UWB-based positioning, and wireless power transfer. He is or has been a TPC Member for numerous IEEE conferences, such as VTC, GLOBECOM, SPAWC, ICC, PIMRC, and WCNC. He was an Elected Member of the Signal Processing for Communications Committee from 2000 to 2005 and the Sensor Array and Multichannel Signal Processing Committee of the Signal Processing Society from 2006 to 2008 and from 2009 to 2011. He was the Chair of the IEEE Benelux Joint Chapter on communications and vehicular technology from 1999 to 2003. He was the Co-Technical Chair of IEEE ICASSP 2006. He served as an Editor of synchronization and equalization for IEEE TRANSACTIONS ON COMMUNICATIONS from 2000 to 2002 and as an Associate Editor for IEEE TRANSACTIONS ON WIRELESS COMMUNICATIONS from 2003 to 2005 and IEEE TRANSACTIONS ON SIGNAL PROCESSING from 2004 to 2006.
\end{IEEEbiography}

\end{document}